\documentclass{iopart}   
\usepackage{graphics}
\usepackage{latexsym}
\usepackage{amsfonts}
\usepackage{epsfig}
\newcommand{\pic}[3]{\begin{figure}
\hspace{2cm}\epsfig{file=becscat_#1.eps,height=#2 cm,angle=270}
\caption[]{\label{becscat:#1} #3}\end{figure}}
 
\newcommand{\be}[1]{\begin{equation}\label{#1}}
\newcommand{\ee}{\end{equation}}   
  
\newcommand{\bea}{\begin{eqnarray}}
\newcommand{\eea}{\end{eqnarray}} 
\newcommand{\eq}[1]{(\ref{#1})}

\renewcommand{\vec}[1]{{\bf #1}}
\begin{document}  
\jl 2
\letter{Atomic scattering from Bose-Einstein condensates}
\author{Ivo H\"aring and Jan--Michael Rost}
\address{Max-Planck-Institute for the Physics of Complex Systems, 
N\"othnitzer Str. 38, D-01187 Dresden, Germany}
\begin{abstract}
\noindent
Elastic scattering probes directly the interaction potential. For 
weakly interacting condensates this potential is given by the 
condensate density. We investigate how the differential and total
cross sections 
reflect the density. In particular, we have determined which signatures the  Thomas Fermi 
approximation leaves in contrast to an exact solution for the 
condensate wave function within the Gross-Pitaevskii theory.
\end{abstract} 
After the pioneering experimental realization of Bose-Einstein
condensation in magnetically trapped atomic gases 
\cite{Anderson(1995),Davis(1995),Bradley(1995)} these
condensates are now routinely generated in many laboratories around
the world.  Consequently the interest is shifting from the production
and structure of the condensates to their dynamics, i.e., their
interaction with matter and radiation.  
One fundamental form of interaction is scattering.  While it is not
entirely clear up to now (see, e.g., \cite{Wynveen(2000)}) how to perform
experimentally a "standard" scattering experiment (that is to generate
projectiles with the appropriate extremely low absolute kinetic
energies) first theoretical studies have explored possible effects
\cite{Idziaszek(1999)}.

The purpose of our work is to investigate the signatures of the target
distribution (i.e.\ the spatial condensate density) in the elastic 
cross section.  In particular, we are interested
in the differences between an exact description of the density in the
framework of the Gross-Pitaevskii theory and the additional Thomas
Fermi approximation which allows one to determine the condensate
density analytically in the limit of a dense and strongly repulsive
atomic gas.

We will study these effects in the first Born
approximation (see also \cite{Idziaszek(1999)}) a natural starting point for
collisionally probing a system at high impact energies (in terms of the
binding energy of the target, i.e. the condensate).
We start from the same setup for  the weakly coupled many-body Bose 
system  as in \cite{Wynveen(2000)} and decompose the Hamiltonian from the beginning 
into 
\be{ham}
H = H_{T}+ H_{P} + U
\ee
where  $H_{T}$ stands for the 
target part containing  the condensate,  $H_{P}$  represents the  projectile 
Hamiltonian  of the impinging atom and $U$ describes its interaction with the target atoms.
 The target Hamiltonian  reads
\be{mbh}
{H_{T}}=\;\int d^3r\hat{\Psi}^\dagger({\vec{r}}) \left(
-\frac{\hbar^2}{2m}\triangle_{{\bf{r}}}+V_{\mbox{ex}} +V_{a}(\vec r)\right) 
\hat{\Psi}({\bf{r}})\,,
\ee
where $\hat\Psi(r)$ is the field operator for destruction of bosons 
at $\vec r$, $V_{\mbox{ex}} = m\omega^2r^2/2$ is the 
harmonic trapping potential. The contact interaction between the 
atoms of mass $m$ is expressed through the s-wave scattering length $a_s$ as
\be{pot}
V_{a}(\vec r) = 
\frac 12\int d^3r' \hat{\Psi}^\dagger({\vec{r'}}) V({\bf{r}},\vec 
r')\hat{\Psi}({\bf{r'}})
\ee
with 
\be{vint}
V(\vec r,\vec r') = 4\pi\hbar^2a_{s}/m \,\delta(\vec r-\vec r').
\ee
For simplicity we consider 
projectile atoms of the same sort as the condensate atoms. 
Hence, the
interaction potential
\be{potU}
U(\vec r_{p})=2V_a(\vec r_p)
\ee
of the condensed atoms with the projectile atom 
is of the same form as the potential $V_{a}(\vec r)$ within the condensate, given 
in \eq{pot}, where $\vec r_{p}$ is the position of the projectile atom.
The factor two has its origin in the single particle nature of the interaction
of the scattering atom with the condensed atoms. 

In the weak coupling limit we may write the field operator as \cite{Bogolubov(1947)}
\be{oper}
\hat\Psi(\vec r) = N_{0}^{1/2}\Psi_{0}(\vec r) + \delta\hat\Psi(\vec r) 
\ee
where $\Psi_{0}$ is a c--number whose squared modulus represents the 
normalized
density of the condensate at position $\vec r$. 
Minimizing the 
grand canonical Hamiltonian $K_T=H_T-\mu N_T$, where $N_T$ is the number
operator of the trapped atoms, 
with respect to the functional $[\Psi_{0}]$ to 
zeroth order in $\delta\hat\Psi(\vec r) $ gives the time-independent 
Gross-Pitaevskii equation for the condensate wave function $\Psi_{0}$
\be{GP}
\left(-\frac{\hbar^2}{2m}\triangle_{{\bf{r}}}+V_{\mbox{ex}} + 
N_{0}4\pi\hbar^2a_{s}/m|\Psi_{0}|^2-\mu\right)\Psi_{0}=0.
\ee
The eigenvalue $\mu$ is the chemical potential.
With (\ref{oper}) and $\Psi_0$ from \eq{GP} one obtains the
elementary quasiparticle excitation modes and energies
  by diagonalizing $K_T$ to second order in $\delta\hat{\Psi}$. This is the
``standard" Bogolubov approximation \cite{Fetter(1972), Griffin(1996)}.

However, in the 
present context we are only interested in elastic collisions. In 
leading order of the number of condensed atoms $N_{0}$  the 
excitations do not contribute to these processes as we will see shortly. 

The asymptotic scattering states are eigenstates of $H_{T}$ and 
$H_{P}$ respectively, i.e. condensate eigenstates  $|\Phi_{i}\rangle$ 
and  $|\Phi_{f}\rangle$ as described above,
and plane waves 
for the projectile atom with asymptotic initial and final momenta $\vec k_{i}$ and $\vec 
k_{f}$, respectively. With equation (\ref{potU}) 
the scattering amplitude in first Born approximation reads  then 
\be{1born}
F(\theta,\phi) = -\frac{m}{2\pi\hbar^2}\int d^3r e^{-i\vec k_{f}\vec 
r}\langle\Phi_{f}|U({\bf{r}})|\Phi_{i}\rangle e^{i\vec k_{i}\vec r}\,
\ee
with the cross section defined as 
\be{cross}
\frac{d\sigma}{d\Omega} = \frac{k_{f}}{k_{i}}|F(\theta,\phi) |^2\, .
\ee

Inserting  \eq{oper} into \eq{1born} and keeping only leading terms in 
$N_{0}$ one obtains the elastic scattering amplitude
\be{elas}
F_{\mbox{el}}(\theta,\phi)  =  -2a_{s}\int d^3r e^{-i\vec q\vec 
r}N_{0}|\Psi_{0}(\vec r)|^2
\ee
where  $\vec q = \vec k_{f}-\vec k_{i}$ is the vector of momentum transfer
and no energy has been exchanged between initial and final state, 
i.e. $\Phi_{i}=\Phi_{f}$ and $|\vec k_{i}| = |\vec k_{f}|\equiv k$.

From \eq{elas} one can see that the elastic cross section probes 
directly the condensate density $|\Psi_{0}|^2$.  For structural properties of 
repulsive atomic gases  with large particle number $N_{0}$ it is often sufficient to 
determine $\Psi_{0}$ in the Thomas-Fermi approximation \cite{Goldman(1981)} as applied, 
e.g., in \cite{Idziaszek(1999)}. 
 
Its justification is most easily seen if one scales 
the unit length and energy in  \eq{GP} 
such that the only remaining parameter appears as an effective mass $\Gamma$.
Before doing so, we separate the relevant radial 
s-wave part  by the substitution $\Psi_{0}(r) = u(r)/r(4\pi)^{-1/2}$ 
and obtain in energy and length units of the 
trapping potential, namely $\hbar\omega$ and $a_{\omega} = 
[\hbar/(m\omega)]^{1/2}$,
\be{GPs}
\left(-\frac{1}{2}\frac{d^2}{dr^2}+\frac{r^2}{2}+ 
\Gamma\frac{u(r)^2}{r^2 }-\frac{\mu}{\hbar\omega}\right)u(r)=0
\ee
where
\be{gamma}
\Gamma = N_{0}a_{s}/a_{\omega}\,.
\ee
Scaling the length in \eq{GPs} according to $r = \Gamma^{1/4}x$, 
dividing \eq{GPs} by $\Gamma^{1/2}$, and defining $\bar u(x) =
u(r)$ leads to
\be{GPs2}
\left(-\frac{1}{2\Gamma}\frac{d^2}{dx^2}+\frac{x^2}{2} + 
\frac{\bar u(x)^2}{x^2}-\bar\mu\right)\bar u(x)=0\,,
\ee
where $\bar\mu = \mu/\hbar\omega\Gamma^{1/2}$.
For large $\Gamma$ the kinetic energy becomes small and is
neglected in the Thomas Fermi approximation (TFA). This leads to the 
solution
\be{TF}
u(r)/r = \left (\mu/\Gamma(1-(r/R)^2)\right)^{1/2}\,,
\ee
where $\mu(\Gamma)= \hbar\omega/2(15\Gamma)^{2/5}$ 
is obtained from the normalization condition 
$\int u^2(r)dr = 1$. It defines simultaneously the cutoff $R(\Gamma) = 
a_{\omega}(15\Gamma)^{1/5}$ which obeys the relation
\be{mu-cutoff}
\mu = \frac{m\omega^2}2R^2.
\ee
 The solution of 
(\ref{GPs}) must be obtained numerically, e.g., by 
imaginary time propagation with the Split Operator Method \cite{Fleck(1976)}.

From the construction it is clear that the TFA is only applicable for
large and positive $\Gamma$ which implies according to \eq{gamma} a
positive scattering length $a_{s}> 0$ and under realistic experimental 
conditions a large particle number $N_{0}\gg 1$. This is illustrated in 
figure \ref{becscat:fig1} 
where the (normalized) solution $\Psi_{0}(r)$  of \eq{GP} is compared 
to the TF-approximate solution \eq{TF}.  Significant differences 
 appear for $\Gamma < 10$, for the order parameter (part a) as well as for the 
 chemical potential (part b). 
However, from  problems involving a simple atom as a target,
one knows that elastic scattering at higher impact energies  probes 
the target distribution to some detail. 
Hence,  it is a priori not clear to what extent the TFA describes 
cross sections well, even in a parameter regime for $\Gamma \gg 1$ 
where it works well for structural properties. 
For total elastic cross sections the best agreement, as expected, is obtained 
at low impact momentum $k$
(figure \ref{becscat:fig2}). The functional dependence  on $k$ is the same 
for the exact solution and the TFA approximation 
for large $k$ and can be derived from the TFA approximation to be
$\sigma(kR\gg 1)\propto (Rk)^{-2}$, where $R = 
a_{\omega}(15\Gamma)^{1/5}$ is the condensate radius in the TFA 
approximation, see \eq{TF}. However, the exact solution and the TFA differ by a 
factor which grows with decreasing $\Gamma$ (figure
\ref{becscat:fig3}a). 

On the other hand, the total cross section for the TFA in the Born approximation 
can be mapped to a single universal curve 
if one scales $\tilde \sigma  = \sigma/\Gamma^2$ and
expresses the momentum in the dimensionless varibale $\tilde k = k R$ where $R$ is 
the cutoff parameter of \eq{mu-cutoff}. The numerical solution does not have
a well defined cutoff parameter (see figure \ref{becscat:fig1}a). However, we
may use \eq{mu-cutoff} to define the cutoff $R_\mu$ from the numerically obtained chemical
potential $\mu$. With this scaling the cross sections from the numerical order parameter
fall onto the same universal curve as the TF cross sections  (\ref{becscat:fig3}b).

Considerably more disagreement between the TFA and the exact solution 
is seen in the differential cross section shown as a function of 
momentum transfer in figure \ref{becscat:fig4}.  One clearly recognizes 
the exponential versus the algebraic decrease in the absolute magnitude
and the different oscillatory behavior.
Both effects can be understood analytically. The "soft edge" of the 
numerically obtained condensate distribution leads to an exponential 
decrease of the cross section with increasing $q$. 
A similar effect occurs
in the photoionization and the Born cross section for scattering 
of metal (jellium) clusters due to a soft edge in the electron 
distribution \cite{Frank(1996),Keller(1997)}.  
On the other hand, the ``hard edge" of the analytical TFA 
distribution \eq{TF} produces via Fourier transform the (unphysical) algebraic 
decrease of the differential cross section. (In the photoionization of clusters this 
would correspond to a box-like electron distribution with a sharp cut-off,
\cite{Frank(1996)}).
For the oscillatory behavior we note that the exact cross section 
oscillates asymptotically for large $q$ and $\Gamma$ with the same frequency
as the TF approximated cross section
given by the condensate radius $R$. 

Overall we conclude that the TFA offers also for scattering problems a reasonably
good approximation as long as $\Gamma$ is not too small. 
Significant deviations from the exact solution occur only for large
 momentum transfer $q$  for which the cross
sections are so small  that they will not be  experimentally accessible.


\section*{References}



\begin{thebibliography}{10}

\bibitem{Anderson(1995)}
Anderson~M~H, Ensher~J~R, Matthews~M~R, Wieman~C~E and Cornell~E~A \newline
1995 {\it Science } {\bf 269 } 198

\bibitem{Davis(1995)}
Davis~K~B, Mewes~M-O, Andrews~M~R, van~Druten~N~J, Durfee~D~S, Kurn~D~M and
Ketterle~W 1995 {\it Phys. Rev. Lett.} {\bf 75} 3969  

\bibitem{Bradley(1995)}
Bradley C C, Sackett C A, Tollett J J and Hulet R G 
1995 {\it Phys. Rev. Lett.} {\bf 75} 1687

\bibitem{Wynveen(2000)}
Wynveen A, Setty A, Howard A, Halley J W and Campell C E
2000 {\it Phys. Rev. A} {\bf 62} 023602

\bibitem{Idziaszek(1999)}
Idziaszek Z, Rz\c{a}\.{z}ewski K and Wilkens M
1999 {\it J. Phys. B} {\bf 32} L205 

\bibitem{Bogolubov(1947)}
Bogolubov N 1947 {\it J. Phys. (USSR)} {\bf 11} 23

\bibitem{Fetter(1972)}
Fetter A F 1972 {\it Ann. Phys. (NY)} {\bf 70} 67

\bibitem{Griffin(1996)}
Griffin A 1996 {\it Phys. Rev. B} {\bf 53} 9341

\bibitem{Goldman(1981)}
Goldman V F and Silvera I F 1981 {\it Phys. Rev. B} {\bf 24} 2870


\bibitem{Fleck(1976)}
Fleck J A, Morris J R, and Feit M D
1976 {\it Appl. Phys.} {\bf 10} 129

\bibitem{Frank(1996)}
Frank O and Rost J M 1996 {\it Z. Phys. D} {\bf 38} 59 
  
\bibitem{Keller(1997)}
Keller S, Engel E, Ast H and Dreizler R M 1997 {\it J. Phys. B} {\bf 30} L703
\end{thebibliography}


\begin{figure}
               \epsfig{file=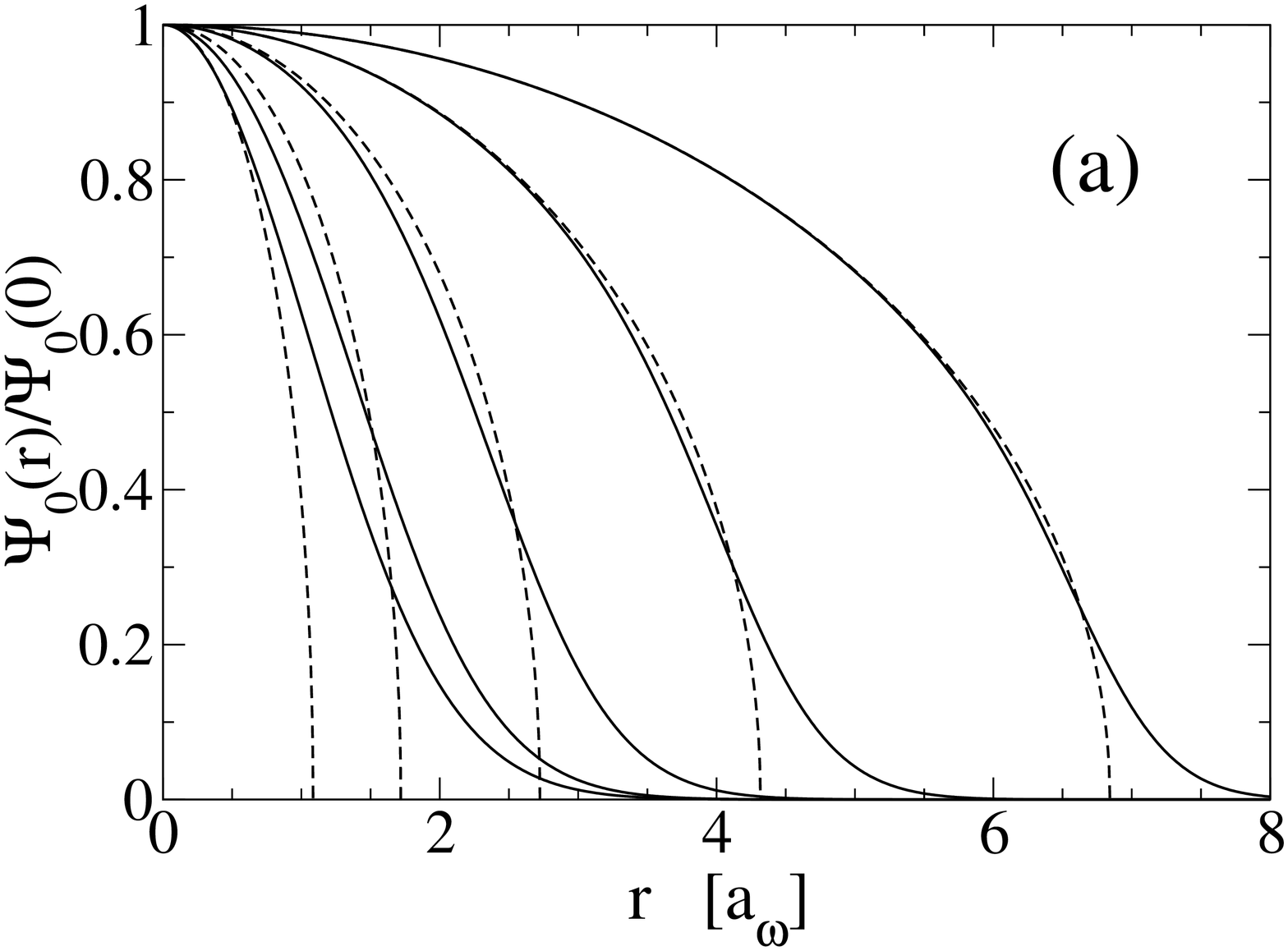,height=7. cm,angle=270}
\hspace{-0.4cm}\epsfig{file=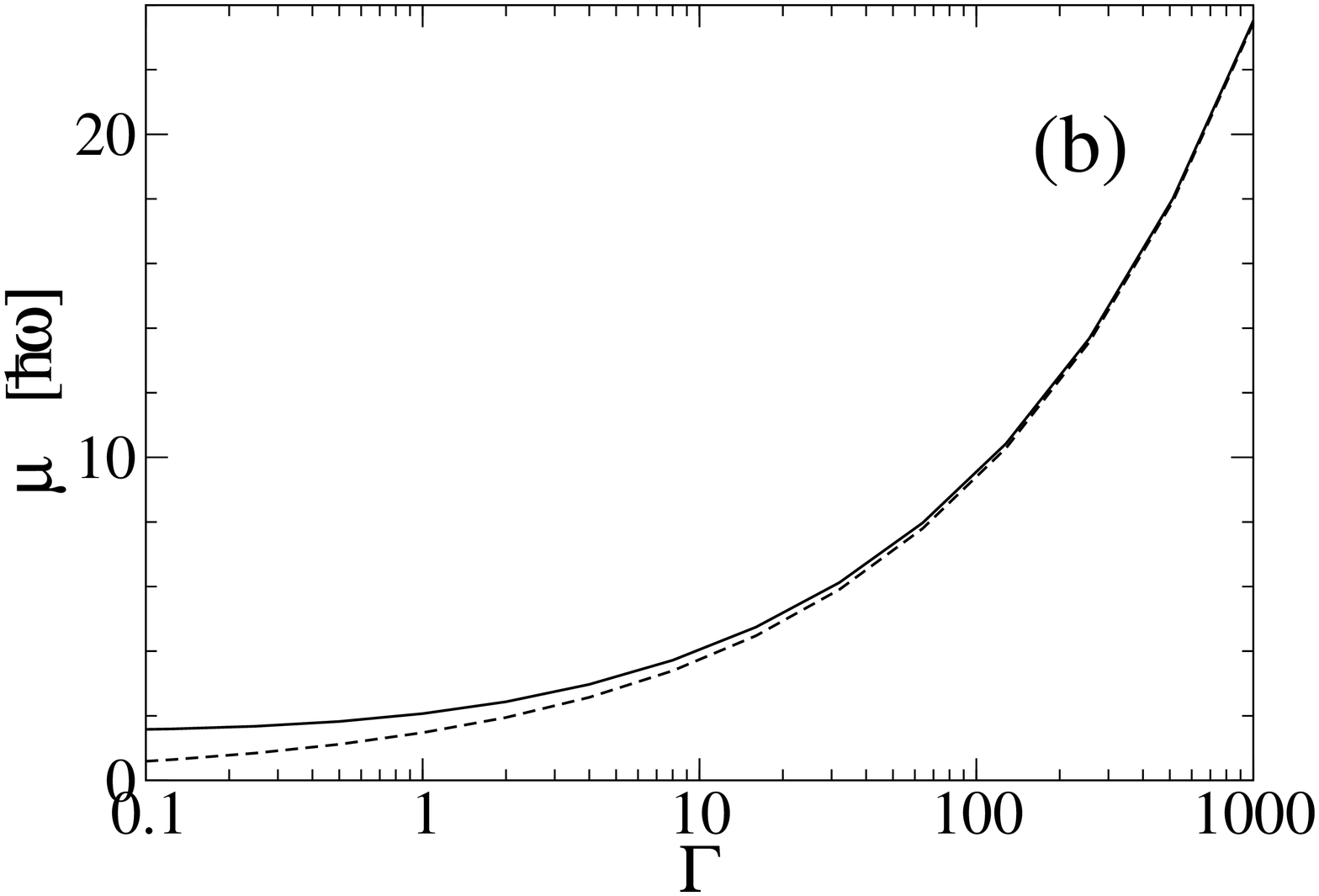,height=7. cm,angle=270}

\caption[]{\label{becscat:fig1}
Properties of the numerical ({\it solid}, \eq{GPs}) and Thomas Fermi 
 ({\it{dashed}}, \eq{TF}) solution of the Gross-Pitaevskii equation.
 Part
(a) presents the order parameter $\psi_0(r)$ normalized to unity at $r=0$  
for $\Gamma = 1/10, 1, 10, 100, 1000$ (from left to right). Part (b) shows 
the chemical potential as a function of $\Gamma$.}
\end{figure}

\pic{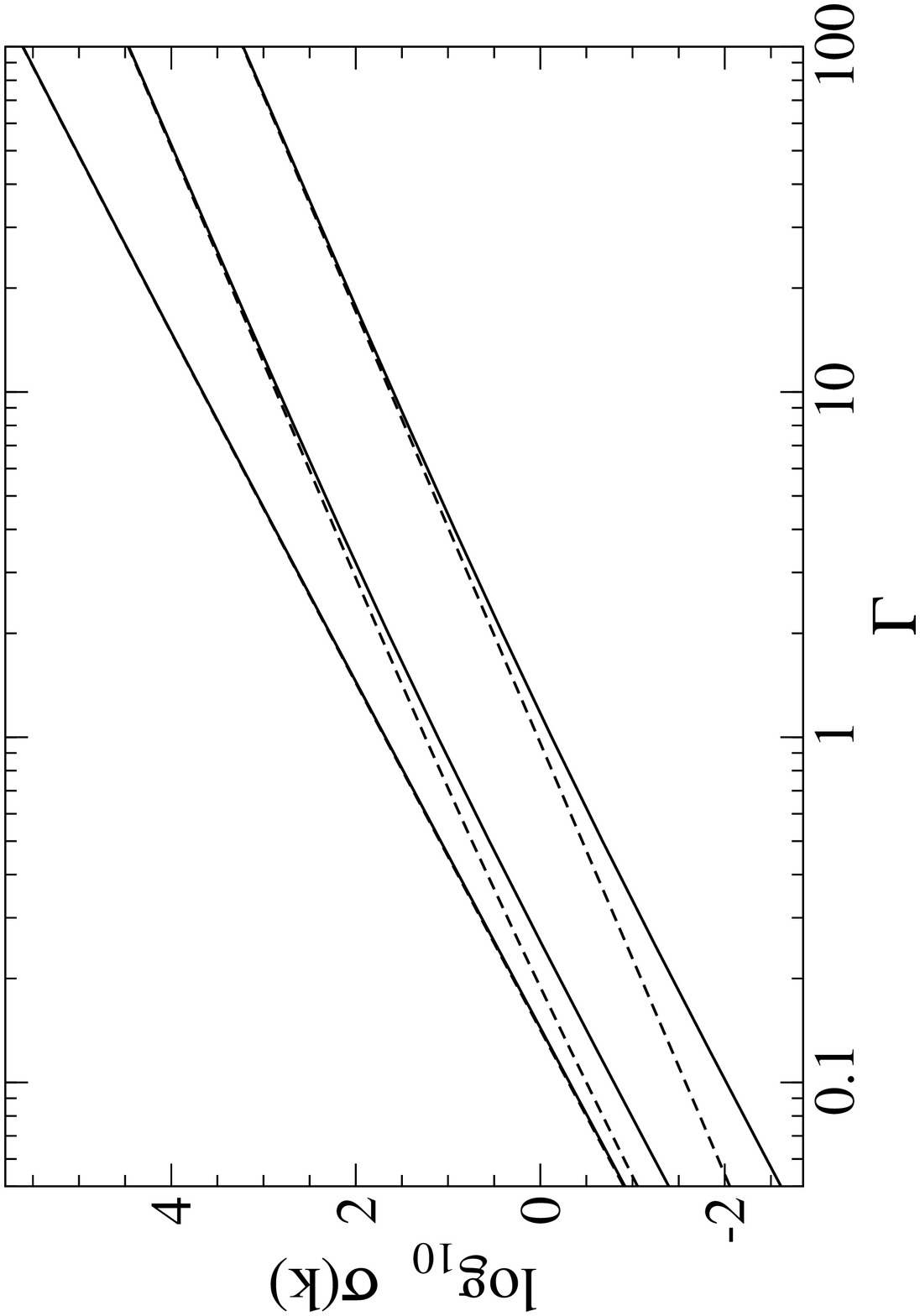}{8}
{
Total elastic cross sections as a function of $\Gamma$ for 
$k = 0.2,1.2, 5$ from top to botton. Coding of the lines as in figure 
\ref{becscat:fig1}.
}

\begin{figure}
                \epsfig{file=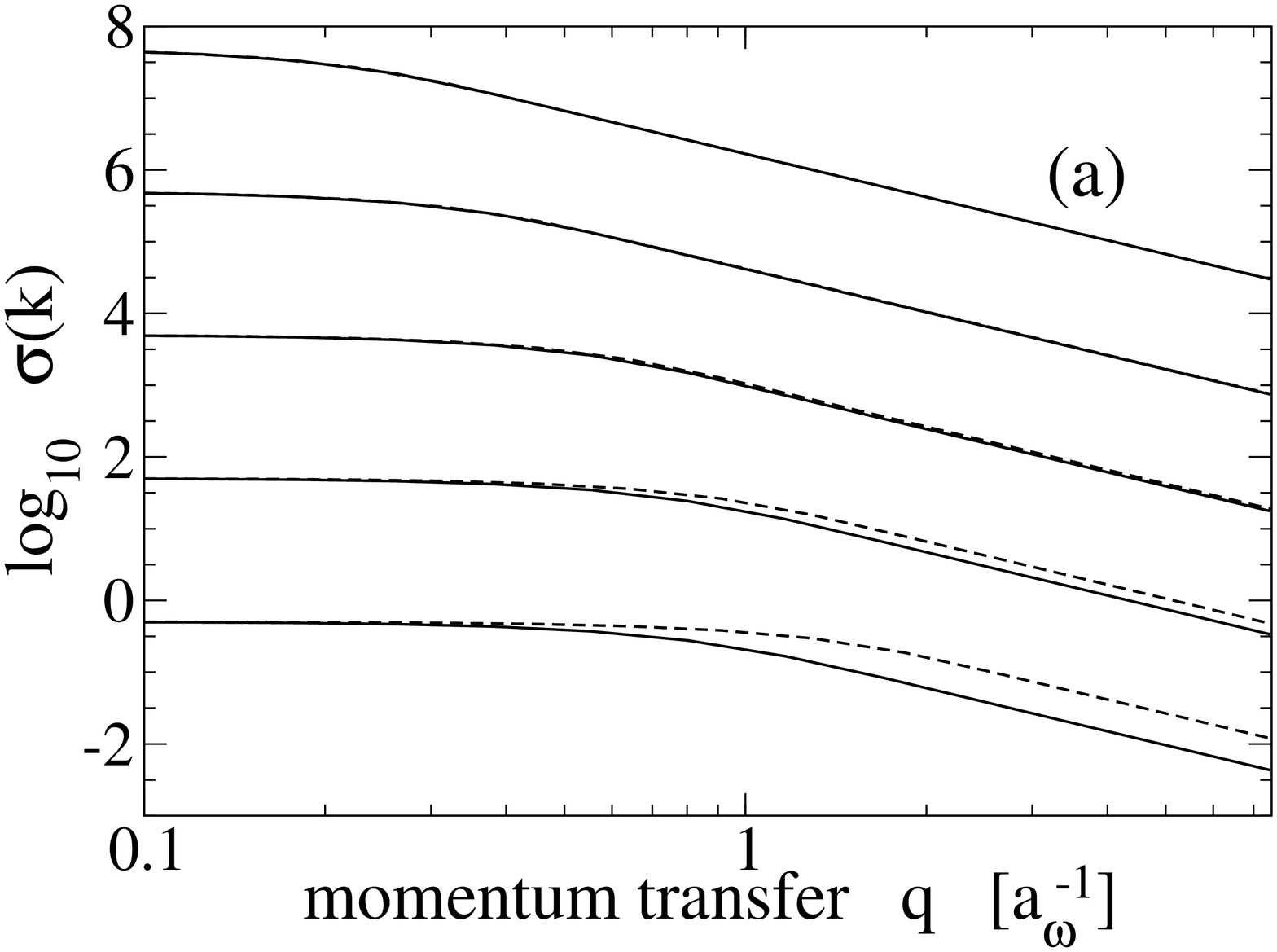,height=7. cm,angle=270}
\hspace{-0.4cm} \epsfig{file=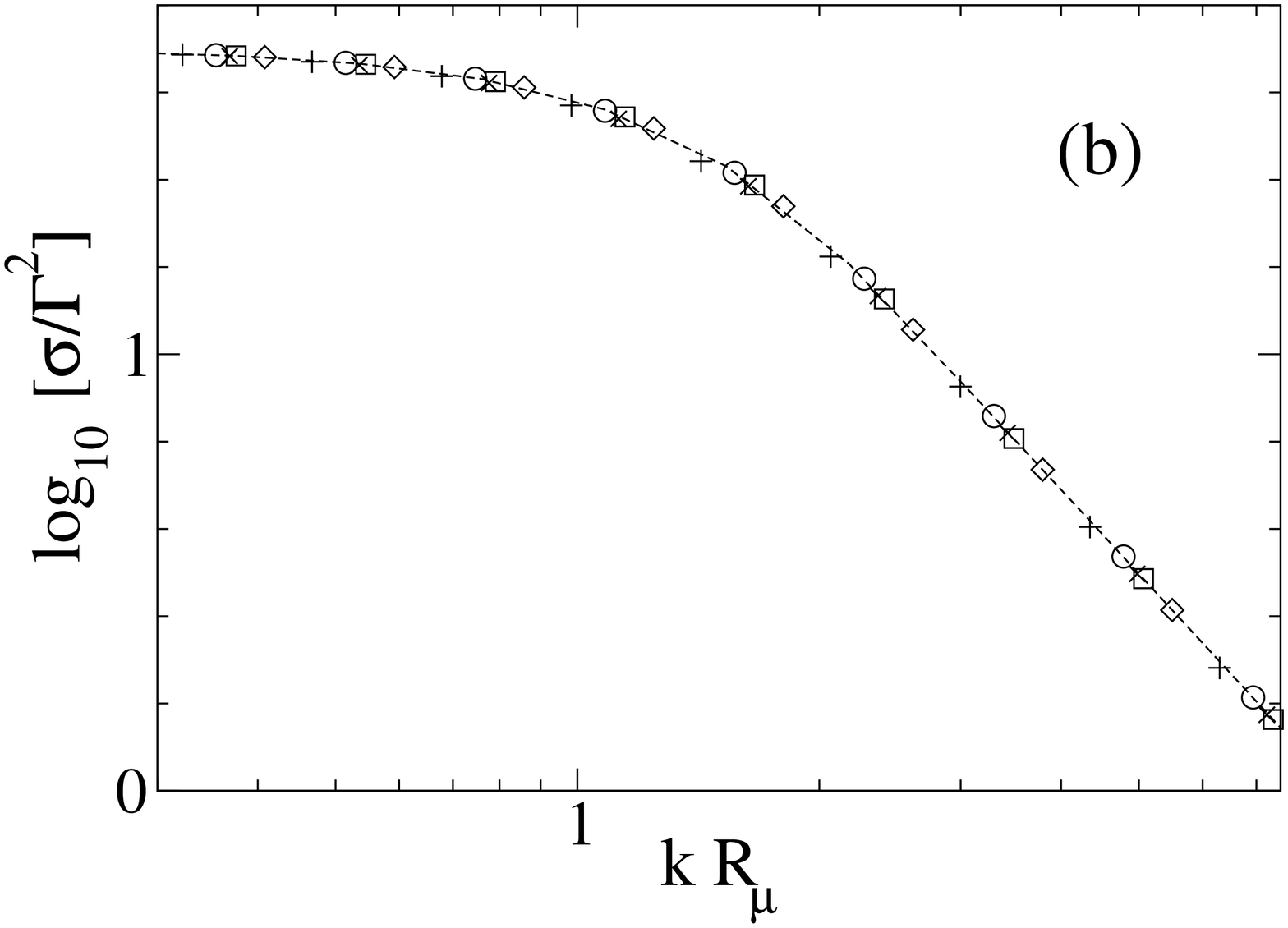,height=7. cm,angle=270}

\caption[]{\label{becscat:fig3} 
(a) Total elastic cross sections for the same  $\Gamma$ values as in figure 
\ref{becscat:fig1}a from top to bottom.
Part (b) shows the cross sections of (a) in scaled coordinates, see text.
The symbols ($+$,$\times$,$\circ$,$\Box$,$\Diamond$) correspond to 
$\Gamma = 1/10,1,10,100,1000$ and the dashed line is the universal TF curve.
}
\end{figure}

\pic{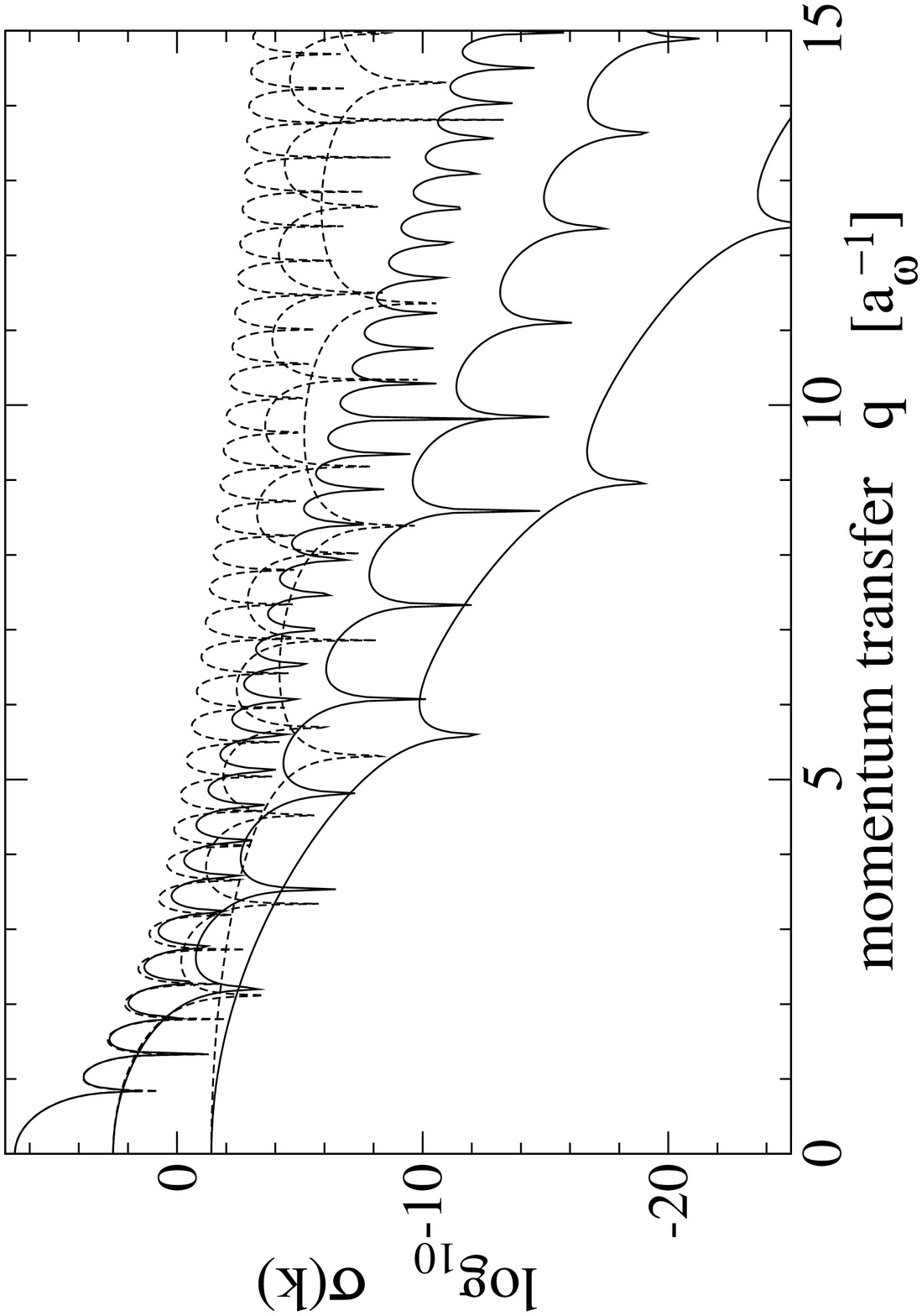}{8}
{
Differential elastic cross sections as a function of momentum transfer   
$q$ for $\Gamma = 1/10, 10, 1000$ from bottom to top
in both the 
group of numerical cross sections ({\it solid}) and analytical cross
sections ({\it dashed}). 
}


\end{document}